\documentclass[twocolumn,showpacs,aps,pra]{revtex4-1}
\usepackage{amssymb}
\usepackage{amsfonts}
\usepackage{dcolumn}
\usepackage{amsmath}
\usepackage{graphicx}
\usepackage{psfrag}
\usepackage{amsmath}
\usepackage{amssymb}
\usepackage{epstopdf}
\usepackage{hyperref}

\newcommand{\be} {\begin{equation}}
\newcommand{\ee} {\end{equation}}

\begin{document}
\title{Entanglement between exciton and mechanical modes via dissipation-induced coupling}
\author{Eyob A. Sete$^{1}$}
\thanks{Present address: Rigetti Quantum Computing, 775 Heinz Ave.,  Berkeley, CA 94710}
\author{H. Eleuch$^2$}

\author{C.H. Raymond Ooi$^3$}
\affiliation{$^1$Department of Electrical and Computer Engineering, University of California, Riverside, California 92521, USA \\
$^2$Department of Physics, McGill University, Montreal, Canada H3A 2T8\\
$^3$Department of Physics, University of Malaya, Kuala Lumpur 50603, Malaysia}

\date{\today}

\begin{abstract}
We analyze the entanglement between two matter modes in a hybrid quantum system consisting of a microcavity, a quantum well, and a mechanical oscillator. Although the exciton mode in the quantum well and the mechanical oscillator are initially uncoupled, their interaction through the microcavity field results in an indirect exciton-mode--mechanical-mode coupling. We show that this coupling is a Fano-Agarwal-type coupling induced by the decay of the exciton and the mechanical modes caused by the leakage of photons through the microcavity to the environment. Using experimental parameters and for slowly varying microcavity field, we show that the generated coupling leads to an exciton-mode--mechanical-mode entanglement. The maximum entanglement is achieved at the avoided level crossing frequency, where the hybridization of the two modes is maximum. The entanglement is also robust against the phonon thermal bath temperature.
\end{abstract}

\maketitle
\section{Introduction}
Hybrid quantum systems consisting of quantum mechanical oscillators have become a platform for many interesting applications of quantum mechanics. In addition to being a tool to understand the quantum to classical transition, e.g., by creating entanglement between mechanical modes, mechanical oscillators have potential applications in quantum information processing. In this regard, there has been a growing effort in exploiting the mechanical degrees of freedom to engineer devices such as a microwave-to-optical (or vise versa) frequency converter \cite{Pal13,And14} and quantum memory \cite{Mcg13,Set15}. Moreover, quantum mechanical oscillator has been used as an interface to transfer a quantum state from an optical cavity to a microwave cavity \cite{Tia12,Cle12}. Interest in merging optomechanical resonators with solid-state systems has been growing \cite{Asp14}; examples include, ultrastrong optomechanical coupling in GaAl vibrating disk resonator \cite{Din10,Din11}, cooling of phonons in a semiconductor membrane \cite{Usa12}, a strong optomechanical coupling in a vertical-cavity resonator \cite{Ang14}, and surface-emitting laser \cite{Fai13}. Coupling a mechanical oscillator to a microcavity that consists of a quantum well has been considered in the context of generating hybrid resonances \cite{Set12} among photons, excitons, and phonons, and studying the optical bistability \cite{Set12,Kyr14}. The physics of photon-exciton coupling has extensively been studied as presented in a recent review \cite{Hui10}.

In this work, we analyze the entanglement between the mechanical mode and the exciton mode in a quantum well placed at the antinode of a microcavity that is formed by distritbuted Brag reflectors (DBRs). Even though the exciton and the mechanical modes are initially uncoupled, their interaction with a common quantized microcavity field results in an indirect coupling. We show that this coupling is a Fano-Agarwal-type \cite{Fan61,Aga76} coupling induced by the decay of the exciton and the mechanical mode caused by the leakage of photons through the microcavity to the environment (Purcell effect \cite{Pur46,Set14}). We analyze the entanglement in the adiabatic regime, where the damping rate of the microcavity exceeds the cavity-exciton coupling strength. A significant amount of entanglement between the exciton and the mechanical modes can be created at the exciton-mechanical mode hybrid resonance frequencies. We find that the maximum entanglement between the two modes is achieved when the exciton and the mechanical modes hybridization is maximum. Surprisingly, the entanglement persists at high temperature of the phonon thermal bath. Our entanglement analysis is based on realistic parameters from a recent experiment \cite{Fai13}.

\begin{figure}[t]
\begin{center}
\includegraphics[width=8cm]{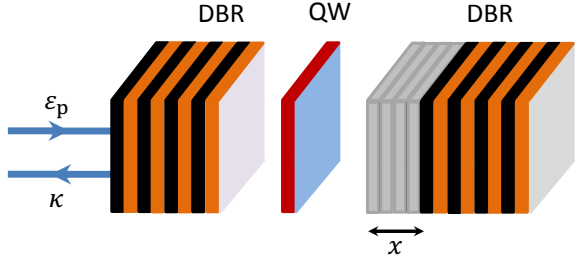}
\caption{Schematic of a microcavity made of two sets of distributed
Bragg reflectors (DBR) mirrors containing a quantum well (QW) and coupled to a
mechanical motion (x) of the mirror. The quantum well is placed at the antinode of the microcavity
so that the exciton-cavity mode coupling will be maximum. The black and the
orange stripes corresponding to GaAs and AlAs layers, respectively. The
microcavity is driven by a pump laser of normalized amplitude $\protect\varepsilon
_{p}$ and has a damping rate $\protect\kappa$. The DBRs are shifted from the equilibrium position due to the radiation pressure force.}
\label{fig1}
\end{center}
\end{figure}
\section{Hamiltonian and Langevin equations}
We consider a microcavity formed by a set of distributed Bragg reflector
mirrors and consists of a quantum well placed at the antinode. The microcavity is
coupled to the mechanical motion of the mirror via radiation pressure force
and to the exciton mode in the quantum well. An exciton in the quantum well can be considered as a quasi-particle resulting from the interaction between
one hole in the valence band and one electron in the conduction band. When the exciton radius is much smaller than the average distance between neighbouring excitons ($\sim n_{\rm ex}^{-1/2}$ with $n_{\rm ex}$ being exciton concentration), we treat the exciton as a composed boson.
In general, in the weak excitation regime, where the density of the excitons is sufficiently low, the interaction between the neighboring excitons due to Coulomb interaction is weak and can be neglected. However, in the moderated driving regime, the interaction between neighbouring excitons becomes strong and nonlinear \cite{Cui00,Tas99,Cui98,Han74a,Han74b,Hau76}, and leads to interesting properties such as squeezing and bistability \cite {Mes99,Ele04,Liu07,Ele10,Set11}. In this paper we will consider the exciton as a composed
boson.

The coupled exciton-optomechanical system is described by the
Hamiltonian
\begin{align}\label{ham}
H& =\omega _{a}a^{\dag }a+\omega _{\mathrm{ex}}b^{\dag }b+\omega _{\mathrm{m}%
}c^{\dag }c +i\varepsilon _{p}(a^{\dag }e^{-i\omega _{p}t}-ae^{i\omega _{p}t}) \notag \\
& -g_{0}a^{\dag }a(c+c^{\dag })+ig(a^{\dag }b-ab^{\dag })+\alpha b^{\dag
}b^{\dag }bb.
\end{align}%
Here the operators $a$, $b$, and $c$ are annihilation operators for a photon in
the microcavity, an exciton in the quantum well, and a phonon in the
mechanical oscillator, respectively. The microcavity is driven by strong drive with frequency $\omega _{p}$;  $\omega _{a}$ and $\omega _{\mathrm{ex}}$ are respectively the bare microcavity and exciton
frequencies. For
the mechanical oscillator, the resonance frequency is $\omega _{\mathrm{m}}$ and
$g_{0}$ is the single-photon
optomechanical coupling; $g$ is the linear exciton-cavity mode
coupling, and $2\alpha=6e^2a_{\rm ex}/\epsilon A$ \cite{Cui00} is the nonlinear coefficient describing the
exciton-exciton scattering due to Coulomb interaction with $e$, $a_{\rm ex}$, $\epsilon$, and $A$ being the electron charge, the exciton Bohr radius, the dielectric constant of the quantum well, and the quantization area, respectively. The strong drive of amplitude $\varepsilon _{p}=\sqrt{\kappa P/\hbar \omega _{p}}$ with $P$ and $\kappa$ being the drive laser power and the microcavity damping rate, respectively, leads to a large steady-state optical filed in the microcavity which increases the occupation numbers in each mode and the optomechanical coupling. The resulting steady-state intracavity amplitude in turn shifts the equilibrium position of the mechanical oscillator through radiation pressure force.

In the Hamiltonian \eqref{ham}, the the first three terms in the first line represent the free energy of the system while the last term describes the coupling of laser drive with the microcavity. In the second line, the first term describes the photon-phonon coupling, the second term represents the linear exciton-photon interaction, and the last term describes the exciton-exciton scattering due to the Coulomb interaction. In a frame rotating with the drive frequency $\omega_p$, the interaction Hamiltonian \eqref{ham} has the form
\begin{align}  \label{V}
V=&-\Delta_{a} a^{\dag}a-\Delta_{\mathrm{ex}}b^{\dag} b+\omega_{\mathrm{m}%
}c^{\dag}c-g_{0}a^{\dag}a (c+c^{\dag})  \notag \\
&+ig(a^{\dag}b-ab^{\dag})+\alpha
b^{\dag}b^{\dag}bb+i\varepsilon_{p}(a^{\dag}-a),
\end{align}
where $\Delta_{a}=\omega_{p}-\omega_{a}$, and $\Delta_{\mathrm{ex}%
}=\omega_{p}-\omega_{\mathrm{ex}}$. Using the interaction Hamiltonian Eq. \eqref{V}, we derive coupled equations for the macroscopic fields, $\bar a$, $\bar b$, and $\bar c$. These equations are obtained by replacing the operators with classical amplitudes in the Heisenberg equations:
\begin{align}
\dot {\bar a}&=-\frac{\kappa}{2}\bar a+i\Delta_{a} \bar a+g\bar b+ig_{0}\bar a (\bar c+\bar{c}^{*})+\varepsilon_{p},\label{a}\\
\dot {\bar b} &=-\frac{\gamma}{2}\bar b+i\Delta_{\rm ex}\bar b-g\bar a -2i\alpha|\bar b|^2\bar b,\label{b}\\
\dot {\bar c}&=-\frac{\gamma_{\rm m}}{2}\bar c-i\omega_{\rm m}\bar c+ig_{0}|\bar a|^2,\label{p}
\end{align}
where $\gamma$ is the exciton spontaneous emission rate, and $\gamma_{\rm m}$ is the damping rate of the mechanical oscillator. The steady state solution to the above equations read
\begin{align}
&\bar c_{s}=\frac{ig_{0}|\bar a_{s}|^2}{\gamma_{\rm m}/2+i\omega_{\rm m}},\\
&\bar a_{s}=-\frac{\gamma/2+i(\Delta_{\rm ex}+2\alpha I_{b})}{g}\bar b_{s},\\
&\bar b_{s}=-\frac{g\varepsilon_{p}}{\kappa\gamma/4+g^2-\tilde \Delta_{a}\tilde \Delta_{\rm ex}+i(\kappa\tilde\Delta_{\rm ex}/2+\gamma\tilde\Delta_{a}/2)},\label{bs}\\
&\tilde \Delta_{a}(I_{b})=\Delta_{a}-\frac{2g_{0}^2\omega_{\rm m}}{\omega_{\rm m}^2+(\gamma_{\rm m}/2)^2}\frac{(\gamma/2)^2+\tilde \Delta_{\rm ex}^2}{g^2}I_{b}^2,\notag\\
&\tilde \Delta_{\rm ex}(I_{b})=\Delta_{\rm ex}+2\alpha I_{b},\notag
\end{align}
where $I_{b}=|\bar b_{s}|^2$ is the steady state exciton number in the quantum well. Note that the Eq. \eqref{bs} yields nonlinear equation for $I_{b}$ in the form
\begin{align}
\frac{I_{b}}{g^2}\left[\left(\frac{\kappa\gamma}{4}+g^2-\tilde \Delta_{a}\tilde \Delta_{\rm ex}\right)^2+\left(\frac{\kappa}{2}\tilde\Delta_{\rm ex}+\frac{\gamma}{2}\tilde\Delta_{a}\right)^2\right]=|\varepsilon_{p}|^2.
\end{align}
The nonlinear equation for $I_b$ is a signature that the exciton number can exhibit bistability \cite{Set12,Kyr14} behaviour for a certain parameter regime. In the following, we discuss
exciton-mechanical mode entanglement in the regime where the system is stable.

The nonlinear quantum Langevin equations can be linearized by writing the
operators as the sum of the steady state classical mean value plus a fluctuating
quantum part: $a=\bar{a}_{s}+\delta a$, $b=\bar{b}_{s}+\delta b$, and $c=\bar{c}_{s}+\delta c$. The linearized
Langevin equations of the fluctuation operators then read
\begin{align}
 \delta \dot{a}&=-\frac{\kappa }{2}\delta a+i\tilde{\Delta}_{a}\delta a+g\delta
b+G(\delta c+\delta c^{\dag })+\sqrt{\kappa }a_{\rm{in}}, \label{da}\\
\delta \dot{b}&=-\frac{\gamma }{2}\delta b+i\tilde{\Delta}_{\rm ex}\delta
b-g\delta a-2i\alpha \bar{b}_{s}^{2}\delta b^{\dag }+\sqrt{\gamma }b_{%
\rm{in}}, \label{db1} \\
\delta \dot{c}&=-\frac{\gamma _{m}}{2}\delta c+i\omega _{\rm{m}}\delta
c+G(\delta a^{\dag }-\delta a)+\sqrt{\gamma _{\rm{m}}}c_{\rm{in}},\label{dc}
\end{align}
where $G=g_{0}\sqrt{\bar{n}_{s}}$ is the many-photon optomechanical coupling
with $\bar{n}_{s}=|\bar{a}_{s}|^{2}$ being the steady state mean photon
number in the microcavity. For simplicity, we have chosen the phase of the
coherent drive such that $\bar{a}_{s}=-i|\bar{a}_{s}|$. Here $a_{\mathrm{in}%
} $, $b_{\mathrm{in}}$, and $c_{\mathrm{in}}$ are the Langevin noise
operators for the microcavity, exciton, and the mechanical modes, respectively. All
noise operators have zero mean, $\langle a_{\mathrm{in}}(\omega )\rangle
=\langle b_{\mathrm{in}}(\omega )\rangle =\langle c_{\mathrm{in}}(\omega
)\rangle =0$. We assume that the microcavity and the quantum well are coupled to
a vacuum reservoir and thus the noise operator are delta-correlated: $\langle
a_{\mathrm{in}}(\omega )a_{\mathrm{in}}^{\dag }(\omega ^{\prime })\rangle
=2\pi \delta (\omega +\omega ^{\prime })$ and $\langle b_{\mathrm{in}%
}(\omega )b_{\mathrm{in}}^{\dag }(\omega ^{\prime })\rangle =2\pi \delta
(\omega +\omega ^{\prime })$. However, the mechanical oscillator is coupled
to a thermal bath and the noise operators have the following nonvanishing
correlation properties in the frequency domain:
$\langle c_{\rm{in}}(\omega )c_{\rm{in}}^{\dag }(\omega ^{\prime
})\rangle  =2\pi (n_{\rm{th}}+1)\delta (\omega +\omega ^{\prime })$, and
$\langle c_{\rm{in}}^{\dag }(\omega )c_{\rm{in}}(\omega ^{\prime
})\rangle  =2\pi n_{\rm{th}}\delta (\omega +\omega ^{\prime })$,
where $n_{\rm{th}}=[\exp (\hbar \omega _{\rm{m}}/k_{B}T)-1]^{-1}$ is
the mean number of thermal phonons with $k_{B}$ being the Boltzmann constant and $T$ the bath temperature.

\section{Exciton-mode--mechanical-mode entanglement}

We next study the entanglement between the exciton and the
mechanical modes in the adiabatic regime, where the microcavity damping rate is larger than the exciton-cavity coupling, $\kappa \gg
g$, the cavity dynamics reaches quasistationary state. We then adiabatically eliminate the cavity mode degrees of freedom by setting $\delta\dot{ a}=0$ in Eq. \eqref{da}. Substituting the resulting equation into Eqs. \eqref{db1} and \eqref{dc}, we
obtain coupled equations for $\delta b$ and $\delta c$ that describe the dynamics of the exciton and the mechanical mode evolutions
\begin{align}
\delta \dot b=&-\frac{\Gamma_{b}}{2}\delta b+i(\tilde{\Delta}_{\rm ex}-\delta\omega_{\rm ex})\delta b-2i\alpha \bar b_{s}^2\delta b^{\dag}\notag\\
&-\frac{1}{2}G_{bc}(\delta c+\delta c^{\dag})-\lambda_b a_{\rm in}+\sqrt{\gamma}b_{\rm in},\label{bb}\\
\delta \dot c=&-\frac{\gamma_{\rm m}}{2}\delta c-i(\omega_{\rm m}+\delta\omega_{\rm m})\delta c
-i\delta\omega_{\rm m}\delta c^{\dag}-\frac{1}{2}G_{bc}\delta b\notag\\
&+\frac{1}{2}G_{bc}^{*}\delta b^{\dag}+\lambda_{c}^{*}a_{\rm in}^{\dag}-\lambda_c a_{\rm in}+\sqrt{\gamma_{\rm m}}c_{\rm in},\label{dcc}
\end{align}
where $\Gamma_{b}=\gamma+\gamma_{b}$ with $\gamma_{b}=4g^2/\kappa[1+(2\tilde \Delta_a/\kappa)^2]$ being the effective relaxation rate of the exciton due to the damping of photons through the microcavity to the environment, also known as the Purcell effect \cite{Pur46,Set14}. Note that the relaxation rate of the exciton is increased by $\gamma_{b}$ as result of interaction with the cavity mode; $\gamma_c=4G^2/\kappa[1+(2\tilde \Delta_a/\kappa)^2]$ is the effective damping rate of the mechanical mode. In contrast to the exciton mode evolution, the cavity-induced relation does not affect the decay term in the Langevin equation for $\delta c$, it does however appear in the noise terms as manifested in Eq. \eqref{dcc}. Note also that the cavity-exciton coupling shifts the exciton frequency by $\delta\omega_{\rm ex}=\gamma_{b}\tilde\Delta_{a}/\kappa$. Similarly, the cavity-mechanical mode coupling gives rise to a shift $\delta\omega_{\rm m}=2\gamma_c\tilde\Delta_a/\kappa$ in the mechanical mode frequency;
$\lambda_{b(c)}=\sqrt{\gamma_{b(c)}}(1+2i\tilde \Delta_{a}/\kappa)[1+(2\tilde \Delta_{a}/\kappa)^2]^{-1/2}$ is the contribution of the cavity-induced dissipation to the noise operator of the exciton (mechanical) mode and finally
\begin{align}\label{Gbc}
G_{bc}=\sqrt{\gamma_{b}\gamma_{c}}\left(1+2i\tilde \Delta_a/\kappa\right)
\end{align}
is the effective exciton-mechanical mode cross coupling. Notice that the cross coupling depends on the effective decay rates $\gamma_{b}$ and $\gamma_{c}$ induced by the photon leakage through the microcavity, which is similar to the Fano-Agarwal effect \cite{Fan61,Aga76}. Dissipation-induced coupling has extensively been explored in quantum optics in creating coherence in three-level atomic systems \cite{Yur06,Set11a,Set11b}. Here we exploit the dissipation-induced coupling to entangle two matter modes: the exciton and the mechanical modes.

To study the entanglement between the exciton and the mechanical modes, it is
more convenient to use the quadrature operators defined by, $\delta
x_{b}=(\delta b^{\dag }+\delta b)/\sqrt{2}$, $\delta y_{b}=i(\delta b^{\dag
}-\delta b)/\sqrt{2}$, $\delta x_{c}=(\delta c^{\dag }+\delta c)/\sqrt{2}$,
and $\delta y_{c}=i(\delta c^{\dag }-\delta c)/\sqrt{2}$ and similar definitions for fluctuation operators $x_{j,\rm in},y_{j,\rm in}$ ($j=a,b$). The equations for
these quadrature operators in matrix form read
\begin{equation}\label{R}
\dot{u}=R u+\mathbf{\eta},
\end{equation}%
where $u=\left(\delta x_{b}, \delta y_{b}, \delta x_{c},\delta y_{c}\right)^{T}$ is vector of quadrature operators and $\mathbf{\eta
}=(F_{x, \rm in}^{b},F_{y, \rm in}^{b},F_{x, \rm in}^{c},F_{y, \rm in}^c)$ with $F_{x, \rm in}^b=-\text{Re}(\lambda_b) x_{a,\mathrm{in}}+\text{Im}(\lambda_b)y_{a,\mathrm{in}}+\sqrt{\gamma }x_{b,\mathrm{in}}$, $F_{y,\rm in}^{b}=-\text{Re}(\lambda_b) y_{a,\mathrm{in}}-\text{Im}(\lambda_b)x_{a,\mathrm{in}}+\sqrt{\gamma }y_{b,\mathrm{in}}$, $F_{x, \rm in}^{c}=\sqrt{\gamma _{\rm m}}x_{c,\mathrm{in}}$, and $F_{y, \rm in}^c=-2\text{Re}(\lambda_c) y_{a,\mathrm{in}}-2\text{Im}(\lambda_c)x_{a,\mathrm{in}}+\sqrt{\gamma _{\rm m}}y_{c,\mathrm{%
in}}$. The diffusion matrix $R$ is given by
\begin{equation*}
R=\left(
\begin{array}{cccc}
-\frac{\Gamma _{b}^{-}}{2} & -\Delta _{\mathrm{ex}}^{+} & \text{Re}(G_{bc}) & 0 \\
\Delta _{\mathrm{ex}}^{-} & -\frac{\Gamma _{b}^{+}}{2} & -\text{Im}(G_{bc}) & 0 \\
0 & 0 & -\frac{\gamma _{\mathrm{m}}}{2} & \omega _{\mathrm{m}} \\
-\text{Im}(G_{bc}) & -\text{Re}(G_{bc}) & -(\omega_{\rm m}+2\delta\omega_{\rm m}) & -\frac{\gamma _{\mathrm{m}}}{2} \\
\end{array}%
\right) ,
\end{equation*}
where $\Gamma _{b}^{\pm }=\Gamma _{b}\pm 4\alpha \text{Im}(\bar{b}_{s}^{2})$
and $\tilde{\Delta}_{\mathrm{ex}}^{\pm }=\tilde{\Delta}_{\mathrm{ex}}-\gamma
_{b}\tilde{\Delta}_{a}/\kappa \pm 2\alpha \text{Re}(\bar{b}_{s}^{2})$.

\begin{figure}[t]
\includegraphics[width=7cm]{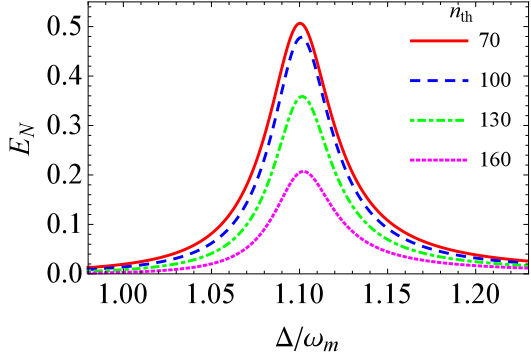}
\caption{Logarithmic negativity $E_{N}$ as a function of the detuning $\Delta=\Delta_{\rm ex}=\Delta_a$ normalized by the bare mechanical resonance frequency $\omega_{\rm m}$ for the input laser power $P=24~\mu \text{W}$, and for various values of the thermal phonon number: $n_{\rm th}=70$~(red solid curve), $100$~ (blue dashed curve), $130$~(green dotdashed curve), and $160$ (magenta dotted curve). Here we use the experimental parameters from a recent experiment \protect\cite{Fai13} $\protect\kappa =1/(5\text{ps}),\protect\gamma =1/(0.5\text{ns}),\gamma %
_{\mathrm{m}}=1/(60\text{ns})$, $g_{0}/2\pi =220~\text{MHz},g/2
\pi =2.4~\text{GHz}$, and $\omega _{\mathrm{m}}/2\pi =20~%
\text{GHz}$, and $\alpha=10^{-9}g$.}
\label{fig2}
\end{figure}

We focus on the steady-state entanglement between
the exciton and the mechanical modes. For this, one needs to find a stable solution for Eq. \eqref{R}, so that it reaches a unique steady state independent of the initial conditions. Since we have assumed $a_{\rm in}$, $b_{\rm in}$, and $c_{\rm in}$ to be zero-mean Gaussian noises and the corresponding equations for fluctuations $\delta x_{j,\rm{in}}$ and $\delta y_{j,\rm{in}
}$ are linearized, the quantum steady state for fluctuations is simply a zero-mean Gaussian state, which is fully characterized by a correlation matrix $V_{ij}=[\langle u_{i}(\infty
)u_{j}(\infty )+u_{j}(\infty )u_{i}(\infty )\rangle ]/2$. The solution to
Eq. \eqref{R} is stable and reaches the steady state when all of the
eigenvalues of $R$ have negative real parts. For all
results presented in this work, the stability has been checked using the nonlinear equation mentioned earlier.
When the system is stable the correlation matrix satisfies Lyapunov equation $RV+VR^{\mathrm{T}}=-D$, where
\begin{equation*}
D=\left(
\begin{array}{cccc}
\frac{\Gamma _{b}}{2} & 0& 0 & 0 \\
0 & \frac{\Gamma _{b}}{2} & 0 & \sqrt{\gamma _{b}\gamma _{c}} \\
0 & 0 & \frac{\gamma _{\mathrm{m}}}{2}(2n_{\mathrm{th}}+1) & 0 \\
0 & \sqrt{\gamma _{b}\gamma _{c}} & 0 & 2\gamma _{c}+\frac{\gamma _{\rm m}}{2}(2n_{\rm th}+1)
\end{array}%
\right)
\end{equation*}
and the elements of the drift matrix $D$ are obtained using the correlations of the noise operators \cite{Set14b} defined earlier. Note that the cavity-induced dissipation terms contribute to the drift matrix. Notably, the off-diagonal element $\sqrt{\gamma_b\gamma_c}=\text{Re}(G_{bc})$ contributes to the correlation between the exciton and the mechanical modes.

In order to quantify the bipartite entanglement, we employ the logarithmic
negativity $E_{N}$, a measure of bipartite entanglement \cite{Vid02,Ple05}. For continuous variables, $E_{N}$ is
defined as
\begin{equation}  \label{LN}
E_{N}=\max [0,-\ln 2\chi],
\end{equation}
where $\chi=2^{-1/2}\left[\sigma-\sqrt{\sigma^2-4\text{det}V}\right]^{1/2}$
is the lowest simplistic eigenvalue of the partial transpose of the $4\times
4$ correlation matrix $V$ with $\sigma=\det V_{A}+\det V_{B}-2\det V_{AB}$ \cite{Ale04}.
Here $V_{A}$ and $V_{B}$ represent the exciton and the mechanical
modes, respectively, while $V_{AB}$ describes the correlation between the two modes.
These matrices are elements of the $2\times2$ block form of the correlation
matrix $V\equiv \left(
\begin{array}{cc}
V_{A} & V_{AB} \\
V_{AB}^{T} & V_{B}
\end{array}
\right)$. The exciton and the mechanical modes are entangled when the logarithmic negativity $E_{N} $ is positive.

\begin{figure}[t]
\includegraphics[width=7cm]{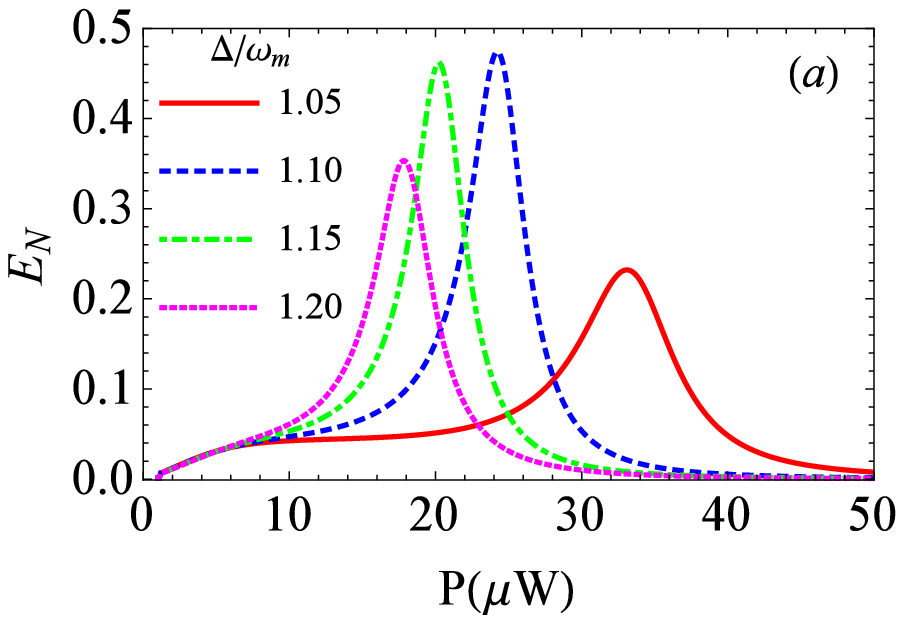}
\includegraphics[width=7cm]{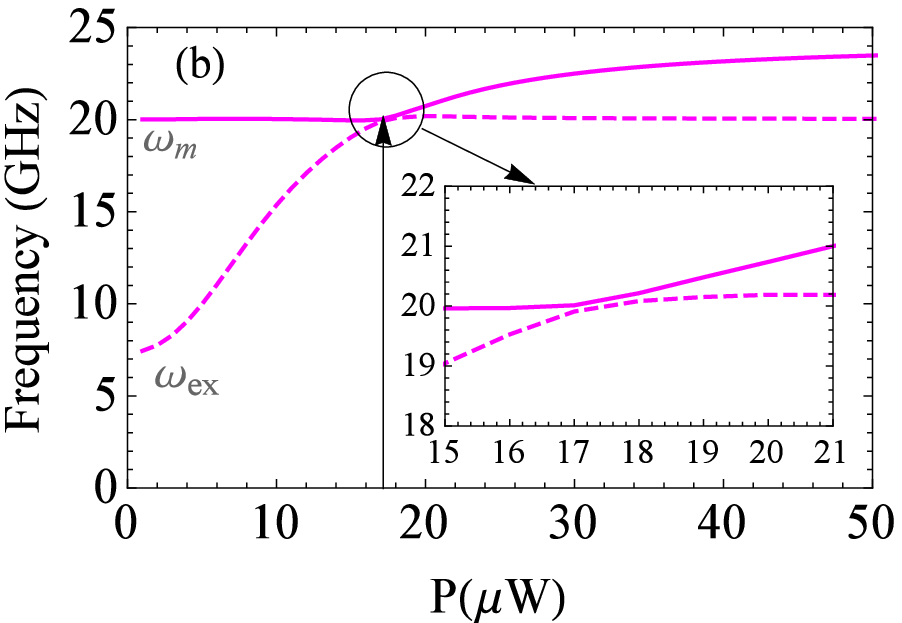}
\caption{(a) Logarithmic negativity as a function of the drive laser power $P$ and for different values of the normalized detuning $\Delta/\omega_{\rm m}=1.05$ (red solid curve), 1.10 (blue dashed curve), 1.15 (green dotdashed curve), and 1.20 (magenta dotted curve). Here we used the thermal photon number $n_{\rm th}=100$. (b) Avoided level crossing between the eigenstates of the exciton-mechanical coupled system for $n_{\rm th}=100$ and $\Delta/\omega_{\rm m}=1.20$. Notice that the maximum entanglement for $\Delta/\omega_{\rm m}=1.20$ in (a) appears at the power ($P\approx 17.8 ~\mu$ W), where the maximum hybridization between the two modes occurs. All the other parameters are as in Fig. \ref{fig2}.}
\label{fig3}
\end{figure}
We numerically studied the exciton-mechanical mode entanglement by exploiting the indirect coupling mediated by the cavity field.
Using realistic parameters from a recent microcavity experiment \cite{Fai13}, we plot in Fig. \ref{fig2} the logarithmic negativity $E_{N}$ as a function of the normalized detuning $\Delta/\omega_{\rm m}$ and for different values of the thermal phonon occupation number, $n_{\rm th}$. Here we assumed the exciton-drive and microcavity-laser detuning are the same, $\Delta_a=\Delta_{\rm ex}=\Delta$.
Figure \ref{fig2} reveals that the exciton and the mechanical modes are strongly entangled, a demonstration of entanglement between two matter modes. The maximum entanglement is achieved at frequency where maximum hybridization between the two modes occurs. The entanglement expectedly decreases when the thermal phonon number is increased; however, it persists up to thermal bath phonon number, $n_{\rm th} \lesssim 200$.

In order to study the dependence of the generated entanglement on the applied input laser power, we plot in Fig. \ref{fig3} the logarithmic negativity versus power for different values of the cavity-laser detuning. As can be seen from this figure, to obtain a maximum entanglement for a given cavity-laser detuning one has to apply a certain laser power strength. Naively, one would expect that an increase in the coupling strength (due to an increase in power) to increase the entanglement. We however find that there exists an optimum amount of power that is needed to obtain the maximum entanglement for the realistic set of parameters \cite{Fai13}. These peaks of the entanglement at different values of the laser power strength and detuning can be explained in terms of the exciton-mechanical mode hybrid resonances. The peaks appear at laser powers where the maximum repulsion between the egienstates of the two modes occur [see, e.g., Fig. \ref{fig3} (b)], indicating that the maximum entanglement is achieved at the maximum of hybridization.
\begin{figure}[t]
\includegraphics[width=7cm]{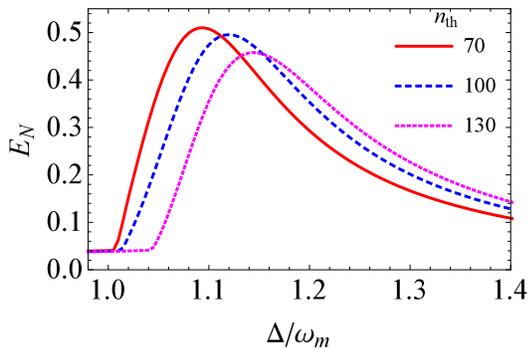}
\caption{Logarithmic negativity as a function of the normalized detuning $\Delta/\omega_{\rm m}$ \textit{optimized} over the input laser power $P$ range: 1-50 $\mu W$ and for different values of the thermal phonon number: $n_{\rm th}=$ 70 (red solid curve), 100 (blue dashed curve), and 130 (magenta dotted curve).}
\label{fig4}
\end{figure}

The optimized entanglement over the input power as a function of detuning and for different values of the thermal phonon number is shown in Fig. \ref{fig4}. The values of the cavity-laser detuning for which the peaks of the entanglement occur shifts when the thermal phonon numbers are varied. This is because the effective coupling [see Eq. \eqref{Gbc}] between the exciton and the mechanical mode depends on the cavity-induced damping rates. These damping rates rely on the number of phonons, thus changing the resonance frequency at which maximum hybridization occurs.

We note that the exciton-mechanical mode entanglement can be detected by measuring the optomechanical entanglement \cite{Pin05,Vit07,Leh13} and the photon-exciton entanglement. From application view point, the generated entangled state has potential in one-way continuous-variable (CV) quantum computation \cite{Su13}. By forming a cluster of entangling gates, it is possible to implement CV quantum computation using our system. The exciton-mechanical mode entanglement might have advantages over that obtained between optical modes \cite{Su13} due to the robustness of the entanglement as well as the availability of semiconductor and micro-electromechanical (MEMS) technologies. The exciton-mechanical mode entanglement also means entanglement with mechanical oscillator or MEMS, a significant progress towards entanglement of macroscopic objects. Achieving entanglement in excitons against its large decoherence is an important step forward as it opens up new possibilities of merging quantum information with existing matured and ubiquitous technologies of semiconducting devices.

\section{conclusion}
In conclusion, we have analyzed the entanglement between two matter modes (exciton and mechanical modes) in a hybrid quantum system consists of a microcavity, a quantum well, and a quantum mechanical oscillator. We have shown that although the exciton and the mechanical modes are initially uncoupled, their interaction with the common microcavity field results in dissipation-induced indirect coupling.  This indirect coupling is responsible for the entanglement between the exciton and the mechanical modes. Maximum entanglement is achieved in the adiabatic regime where the microcavity damping rate is larger than the coupling strengths and when the two modes form a complete hybridization. Recent successful experiments \cite{Din11,Usa12,Fai13} in coupling mechanical systems with microcavity pave the way for the realization of the proposed entanglement generation between exciton and the mechanical modes via dissipation-induced coupling.

\begin{acknowledgements}
EAS acknowledges support from the Office of the Director of National Intelligence (ODNI), Intelligence
Advanced Research Projects Activity (IARPA), through the Army
Research Office Grant No. W911NF-10-1-0334. All statements of fact,
opinion or conclusions contained herein are those of the authors and
should not be construed as representing the official views or
policies of IARPA, the ODNI, or the U.S. Government. He also
acknowledge support from the ARO MURI Grant No. W911NF-11-1-0268. C. H. R. Ooi acknowledges
support from the Ministry of Higher Education of Malaysia
through the High Impact Research MoE Grant UM.C/625/1/
HIR/MoE/CHAN/04.
\end{acknowledgements}

\end{document}